\documentclass[aps,prb,reprint,twocolumn,superscriptaddress,showpacs,floatfix,longbibliography]{revtex4-1}

\usepackage{amsmath}
\usepackage{color}
\usepackage{mathrsfs}
\usepackage{dcolumn}
\usepackage{bm}
\usepackage{multirow}
\usepackage{graphicx}
\usepackage{rotating}
\usepackage{layout}
\usepackage[version=3]{mhchem}
\usepackage{braket}
\usepackage{longtable}
\usepackage{setspace}

\allowdisplaybreaks
\usepackage{xfrac}

\usepackage[colorlinks,citecolor=blue,urlcolor=blue,bookmarks=false,hypertexnames=true]{hyperref}

\begin{document}
\title{Note on a rigorous derivation of self-consistent double-hybrid functional theory via generalized Kohn-Sham theory and cumulant approximation}
\date{\today}
\author{Lan Nguyen Tran}
\email{tnlan@hcmus.edu.vn}
\affiliation{University of Science, Vietnam National University, Ho Chi Minh City 700000, Vietnam}
\affiliation{Vietnam National University, Ho Chi Minh City 700000, Vietnam}
\date{\today}

\begin{abstract}
In this short note, we present a rigorous derivation of the one-body double-hybrid density functional (OBDHF) theory, a self-consistent double-hybrid density functional framework that unifies the generalized Kohn-Sham (GKS) formalism with one-body Møller-Plesset second-order perturbation (OBMP2) theory. Conventional double-hybrid density functionals suffer from a fundamental theoretical inconsistency arising from the non-self-consistent treatment of the perturbative MP2 correlation, in which the orbitals entering the correlation energy expression are not variationally optimized with respect to the full double-hybrid energy functional. To address this deficiency, we construct a model energy functional as a linear combination of semilocal density functional approximation XC, a fraction $\alpha_x$ of exact Hartree-Fock (HF) exchange, and a fraction $\alpha_c$ of OBMP2 correlation. By virtue of the one-body operator structure of OBMP2, the perturbative correlation contribution is embedded directly and self-consistently into the GKS effective Hamiltonian, without recourse to the optimized effective potential (OEP) construction or perturbative orbital relaxation corrections. Through functional differentiation of the total OBDHF energy with respect to the orbitals, we derive the OBDHF effective Hamiltonian and the associated self-consistent field equations in a rigorous and transparent manner. This formulation provides a theoretically well-founded and practically tractable pathway to fully self-consistent double-hybrid density functional theory calculations within the GKS framework, resolving the self-consistency problem inherent in conventional double-hybrid functionals. 
\end{abstract}
\maketitle

\section{Introduction}
Density functional theory (DFT), established on the foundational theorems of Hohenberg and Kohn~\cite{Hohenberg1964} and the practical self-consistent field framework of Kohn and Sham~\cite{Kohn1965}, has become the workhorse
of modern electronic structure calculations in chemistry, physics, and materials science. Its favorable balance between computational cost and accuracy has enabled the study of ground-state properties of systems ranging
from small molecules to extended solids. Nevertheless, the predictive power of Kohn-Sham DFT is fundamentally contingent upon the quality of the exchange-correlation (XC) functional, whose exact form remains unknown and
must be approximated in practice~\cite{Burke2012, Jones2015}.

Standard semilocal approximations, including the local density approximation (LDA)~\cite{Kohn1965, Perdew1992} and the generalized gradient approximation (GGA)~\cite{Perdew1996, Becke1988}, capture a broad range of physical and chemical phenomena, yet suffer from well-documented systematic deficiencies:
self-interaction error~\cite{Perdew1981}, delocalization
error~\cite{Cohen2008}, and an inadequate description of London dispersion and long-range dynamical correlation effects~\cite{Klimes2012}.
These limitations motivate the continued development of more sophisticated XC approximations along the  Jacob's ladder of DFT proposed by Perdew and Schmidt~\cite{Perdew2001}.

A significant advance along this ladder was the introduction of hybrid functionals by Becke~\cite{Becke1993a}, which incorporate a fraction of
nonlocal HF exchange into the XC functional. Within the generalized Kohn-Sham (GKS) framework formalized by Seidl, G\"orling, Vogl,
Majewski, and Levy~\cite{Seidl1996}, the inclusion of nonlocal, orbital-dependent operators in the effective potential is rigorously justified. Hybrid functionals such as B3LYP~\cite{Becke1993a, Lee1988, Stephens1994}, PBE0~\cite{Adamo1999, Ernzerhof1999}, and HSE06~\cite{Heyd2003, Heyd2006} have demonstrated substantially improved performance for thermochemistry,
reaction barriers, band gaps, and molecular geometries compared to their semilocal counterparts.

A further rung of Jacob's ladder is double-hybrid (DH) density
functionals, originally proposed by Grimme~\cite{Grimme2006}, which augment
the hybrid exchange with a perturbative second-order M\o{}ller-Plesset (MP2)
correlation contribution~\cite{Moller1934}. Subsequent developments have
produced a rich family of DH functionals, including B2-PLYP~\cite{Grimme2006},
mPW2-PLYP~\cite{Schwabe2006}, XYG3~\cite{Zhang2009}, PBE0-DH~\cite{Sharkas2011},
PBE-QIDH~\cite{Bremond2011}, and $\omega$B97X-2~\cite{Chai2009}, among many
others~\cite{Goerigk2014, Sancho-Garcia2013, Bremond2016}. These functionals
have demonstrated state-of-the-art accuracy across diverse benchmark sets
including thermochemistry, kinetics, non-covalent interactions, and excited states~\cite{Goerigk2017, Mardirossian2017}.

Despite their remarkable accuracy, conventional DH functionals suffer from a
fundamental theoretical inconsistency: the MP2 correlation energy is evaluated
as a non-self-consistent, \textit{post}-SCF correction applied to orbitals
optimized under the hybrid XC functional without the MP2 contribution. As a
consequence, the orbitals entering the perturbative correlation expression are
not variationally optimized with respect to the full DH energy functional.
This non-self-consistency implies that the one-particle density matrix, and
hence physical observables sensitive to it, such as the electron density,
dipole moment, and response properties, are not fully consistent with the DH
energy expression~\cite{Sharkas2011, Toulouse2011, Fromager2011}. Achieving
a self-consistent treatment of DH functionals therefore represents
an important open challenge in the development of orbital-dependent density
functional approximations.

One rigorous route to self-consistent inclusion of perturbative correlation
within the KS framework is provided by the optimized effective potential (OEP)
method~\cite{Sharp1953, Talman1976, Kummel2008}, which constructs a local
multiplicative XC potential corresponding to an orbital-dependent functional.
However, the OEP procedure is computationally demanding and numerically ill-conditioned in finite basis sets, limiting its practical
applicability~\cite{Hirata2001, Bartlett2005, Grabowski2014}. Alternatively,
within the GKS framework, one may directly include the orbital-dependent
correlation contribution through a nonlocal effective operator, bypassing the
need for the OEP altogether~\cite{Seidl1996, Kummel2008}.

A particularly natural framework for wrapping MP2-level correlation
self-consistently into a GKS orbital equation is offered by the one-body
M\o{}ller-Plesset second-order perturbation (OBMP2) theory, recently
developed by the author~\cite{OBMP2-JCP2013,OBMP2-JPCA2021,OBMP2-PCCP2022,OBMP2-JPCA2023,OBMP2-JPCA2024,OBMP2-JCP2025}. OBMP2 formulates
the MP2 dynamic correlation effects as an effective one-body correlated Fock
operator, obtained through a unitary canonical transformation of the molecular
Hamiltonian~\cite{CT-JCP2006,CT-JCP2007,CT-ACP2007,CT-JCP2009,CT-JCP2010,CT-IRPC2010} followed by the cummulant approximation to reduce many-body operators to one-body operators. Because OBMP2 yields a
one-body correlated potential operator rather than a two-body perturbative energy, it is uniquely
suited for direct incorporation into the GKS effective Hamiltonian without
recourse to response equations, orbital gradient corrections, or the OEP construction.

In this note, we present the development of self-consistent \textit{one-body double-hybrid} (OBDHF)
density functional theory that combines the GKS framework with OBMP2 theory. The central contribution of
this note is a rigorous derivation of the OBDHF orbital equation, in which
both a fraction $\alpha_x$ of exact HF exchange and a fraction $\alpha_c$ of
OBMP2 correlation are embedded self-consistently within the GKS model functional and the corresponding orbital optimization. The model functional is constructed as a linear combination of semilocal DFA XC, exact exchange, and OBMP2 correlation. By taking the functional derivative of the total energy with respect to the orbitals, we derive the OBDHF effective Hamiltonian and the associated self-consistent field (SCF)
equations. This formulation provides a theoretically
rigorous and practically tractable pathway to fully self-consistent double-hybrid DFT calculations within the GKS framework, without invoking the OEP or perturbative orbital relaxation corrections.

\section{Theory}
\subsection{Generalized Kohn-Sham theory}
The Kohn-Sham (KS) formalism maps the original interacting electron problem to a fictitious \textit{non-interacting} system. This mapping is constructed such that the fictitious system retains the exact same ground-state electron density, $n(\mathbf{r})$, as the real physical system. The core of this approach is the Kohn-Sham equation, which determines the single-particle orbitals needed to construct the density:
\begin{equation}
    \left( -\frac{1}{2}\nabla^2 + V_{KS}([n]; \mathbf{r}) \right) \phi_i(\mathbf{r}) = \varepsilon_i \phi_i(\mathbf{r})
\end{equation}
Where, $\varepsilon_i$ are the Kohn-Sham eigenvalues, $\phi_i(\mathbf{r})$ are the Kohn-Sham orbitals, and $V_{KS}([n]; \mathbf{r})$ is the effective multiplicative potential, which is a functional of the electron density at each point in space. It is typically decomposed into three specific terms:
\begin{equation}
    V_{KS}([n]; \mathbf{r}) = V_{ext}(\mathbf{r}) + V_{H}([n]; \mathbf{r}) + V_{xc}([n]; \mathbf{r})
\end{equation}
Where
\begin{itemize}
    \item $V_{ext}$ is the external potential representing the static ionic potential of the system (the attraction between electrons and nuclei).
    \item $V_{H}$) is the Hartree potential accounting for the classical electron-electron repulsion. It is defined as:
        \begin{equation}
            V_{H}([n]; \mathbf{r}) = \int \frac{n(\mathbf{r}')}{|\mathbf{r} - \mathbf{r}'|} d\mathbf{r}'
        \end{equation}
    \item $V_{xc}$ is the XC potential accounting for all many-body quantum effects that go beyond classical repulsion. It is a functional of the density, though the exact form is typically unknown and must be approximated.
\end{itemize}

In GKS theory, we define an energy functional of the Slater determinant, denoted as $S[\Phi]$ or equivalently as a functional of the orbitals $S[\{\phi_j\}]$. An associated energy density functional, $F_S[n]$, is defined by minimizing $S$ over all Slater determinants that yield the specific density $n(\mathbf{r})$:

\begin{equation}
    F_S[n] \equiv \min_{\{\phi_j\} \to n(\mathbf{r})} S[\{\phi_j\}].
\end{equation}
Here, the minimizing orbitals $\{\phi_j\}$ play a role analogous to standard KS orbitals.

Using the Hohenberg-Kohn theorem, the universal functional $F_{HK}[n]$ is decomposed into the model energy $F_S[n]$ and a ``remainder energy'' functional $R_S[n]$:
\begin{equation}
    R_S[n] = F_{HK}[n] - F_S[n].
    \label{eq:R_s}
\end{equation}
Consequently, the total energy of the system can be expressed as:
\begin{equation}
    E_{tot} = \min_{n} \left( \int V_{ext}(\mathbf{r})n(\mathbf{r}) d\mathbf{r} + F_S[n] + R_S[n] \right).
\end{equation}

By minimizing the energy with respect to the underlying orbitals (subject to the constraint that they integrate to the density), one derives the generalized KS equation:
\begin{equation}
    \left( \hat{O}_S[\{\phi_j\}] + V_{ext}(\mathbf{r}) + V_R(\mathbf{r}) \right) \phi_i(\mathbf{r}) = \varepsilon_i \phi_i(\mathbf{r}),
    \label{eq:ks}
\end{equation}
where:
\begin{itemize}
    \item $V_R(\mathbf{r})$: A \textbf{multiplicative} ``remainder potential,'' defined as the functional derivative of the remainder energy:
    \begin{equation}
        V_R(\mathbf{r}) \equiv \frac{\delta R_S[n]}{\delta n(\mathbf{r})}.
    \end{equation}
    \item $\hat{O}_S[\{\phi_j\}]$: A generally \textbf{non-multiplicative} operator derived from the model functional $S$. This operator obviously depends on the choice of $S[.]$, but does not depend on $V_{ext}(\mathbf{r})$ and $V_R(\mathbf{r})$.
\end{itemize}


\subsection{One-body M{\o}ller-Plesset second-order perturbation (OBMP2) theory}
Let us recap the OBMP2 theory presented in Refs.~\citenum{OBMP2-JCP2013,OBMP2-JPCA2021,OBMP2-PCCP2022,OBMP2-JPCA2023,OBMP2-JPCA2024,OBMP2-JCP2025}. For convenience, we use Einstein's convention to present summations over repeated indices. The OBMP2 approach was derived through the canonical transformation \cite{CT-JCP2006,CT-JCP2007,CT-ACP2007,CT-JCP2009,CT-JCP2010,CT-IRPC2010}, in which an effective Hamiltonian that includes dynamic correlation effects is achieved by a similarity transformation of the molecular Hamiltonian $\widehat{H}$ using a unitary operator $e^{\widehat{A}}$ with the anti-Hermitian operator $\widehat{A} = -\widehat{A}^\dagger$:
\begin{align}
\widehat{\bar{H}} = e^{-\widehat{A}} \widehat{H} e^{\widehat{A}},
\label{Hamiltonian:ct}
\end{align}
with the molecular Hamiltonian as
\begin{align}
  \widehat{H} =  \sum_{pq}h^{p}_{q} \widehat{a}_{p}^{q} + \tfrac{1}{2}\sum_{pqrs}g^{p r}_{q s}\widehat{a}_{p r}^{q s}\label{eq:h1}.
\end{align}
Here, $\left\{p, q, r, \ldots \right\}$ indices referring to general ($all$) spin orbitals. One- and two-body second-quantized operators $\widehat{a}_p^q$ and $\widehat{a}_{pq}^{rs}$ are given by $\widehat{a}_p^q = \widehat{a}^\dagger_p\widehat{a}_q$ and $\widehat{a}_{pq}^{rs} = \widehat{a}^\dagger_p\widehat{a}^\dagger_q\widehat{a}_s\widehat{a}_r$. One- and two-electron integrals, $h_{pq}$ and $v_{pq}^{rs}$, are respectively defined as
\begin{align*}
    h_{pq} = \int \phi_p^*(\vec{r_1})\left(-\tfrac{1}{2}\nabla^2-\sum_{I=1}^{M}\frac{Z_{I}}{r_{1I}}\right)\phi_q(\vec{r_1 }) d\vec{r_1 }, \\
    v_{pq}^{rs}=\int \phi_p^*(\vec{r_1 }) \phi_q^*(\vec{r_2})\frac{1}{r_{12}} \phi_s(\vec{r_2}) \phi_r(\vec{r_1}) d\vec{r_1 } d\vec{r_2},
\end{align*}
where $Z_I$ is the nuclear charge of atom $I$, and $r_{1I} = \left|\vec{r_1}-\vec{R_I}\right|$ and $r_{12} = \left|\vec{r_1}-\vec{r_2}\right|$.

In OBMP2, the anti-Hermitian excited operator $\widehat{A}$ includes only double excitations. 
\begin{align}
  \widehat{A} = \widehat{A}_\text{D} = \tfrac{1}{2} \sum_{ij}^{occ} \sum_{ab}^{vir} T_{ij}^{ab}(\widehat{a}_{ab}^{ij} - \widehat{a}_{ij}^{ab}) \,, \label{eq:op1}
\end{align}
with the MP2 amplitude 
\begin{align}
  T_{i j}^{a b} =  \frac{g_{i j}^{a b} } { \epsilon_{i} + \epsilon_{j} - \epsilon_{a} - \epsilon_{b} } \,, \label{eq:amp}
\end{align}
where $\left\{i, j, k, \ldots \right\}$ indices refer to occupied ($occ$) spin orbitals and
$\left\{a, b, c, \ldots \right\}$ indices refer to virtual ($vir$) spin orbitals; $\epsilon_{i}$ is the orbital energy of the spin-orbital $i$. Using the Baker–Campbell–Hausdorff transformation and several approximations as reported in Refs.~\citenum{OBMP2-JCP2013} and \citenum{OBMP2-JPCA2021}, the OBMP2 Hamiltonian is defined as
\begin{align}
  \widehat{H}_\text{OBMP2} = \widehat{H}_\text{HF} + \left[\widehat{H},\widehat{A}\right]_1 + \tfrac{1}{2}\left[\left[\widehat{F},\widehat{A}\right],\widehat{A}\right]_1.
 \label{eq:h2}
\end{align}
with
\begin{align}
  \widehat{H}_\text{HF} &= \widehat{F} + C = \widehat{h} + \widehat{v}_{\text{HF}} +C \label{eq:h1hf}
\end{align}
Where $\widehat{h}$ is the one-electron Hamiltonian defined in Eq~\ref{eq:h1}. 
The HF potential $\widehat{v}^{\text{HF}}$ and the constant $C$ is given as:
\begin{align}
    \hat{v}_{\text{HF}} &= 
    \sum_{pq}^{\text{all}} \left[\sum_{i}^{\text{occ}} 
\left( g^{pi}_{qi} - g^{pi}_{iq} \right)\right] \hat{a}^{p}_{q} \\
    C &=  \,\, \sum_{ij}^{occ} \left(g^{ij}_{ji} - g_{ij}^{ij} \right) \,. 
\end{align}
In Eq.\ref{eq:h2}, commutators with the subscription 1, $[\ldots]_1$, involve one-body operators and constants that are reduced from many-body operators using the cumulant approximation\cite{cumulant-JCP1997,cumulant-PRA1998,cumulant-CPL1998,cumulant-JCP1999}. After some derivation, we arrive at the OBMP2 Hamiltonian as
\begin{align}
  \widehat{H}_\text{OBMP2} = & \,\, \widehat{H}_\text{HF} + \widehat{v}_\text{OBMP2} \label{eq:h4}
\end{align}
where $\widehat{v}_\text{OBMP2}$ is a correlated potential composing of one-body operators. The working expression is given as
\begin{align}
\widehat{v}_{\text{OBMP2}} = &  \overline{T}_{i j}^{a b} \left[ f_{a}^{i} \,\widehat{\Omega}\left( \widehat{a}_{j}^{b} \right) 
  + g_{a b}^{i p} \,\widehat{\Omega} \left( \widehat{a}_{j}^{p} \right) - g^{a q}_{i j} \,\widehat{\Omega} \left( \widehat{a}^{b}_{q} \right) \right] \nonumber \\  &- 2 \overline{T}_{i j}^{a b}g^{i j}_{a b} 
   + \,f_{a}^{i}\overline{T}_{i j}^{a b}\overline{T}_{j k}^{b c} \,\widehat{\Omega} \left(\widehat{a}_{c}^{k} \right) \nonumber \\ 
     &+  f_{c}^{a}T_{i j}^{a b}\overline{T}_{i l}^{c b} \,\widehat{\Omega} \left(\widehat{a}^{l}_{j} \right) + f_{c}^{a}T_{i j}^{a b}\overline{T}_{k j}^{c b} \,\widehat{\Omega} \left(\widehat{a}^{k}_{i} \right) \nonumber \\ 
     &-  f^{k}_{i}T_{i j}^{a b}\overline{T}_{k l}^{a b} \,\widehat{\Omega} \left(\widehat{a}_{l}^{j} \right)
     -  f^{p}_{i}T_{i j}^{a b}\overline{T}_{k j}^{a b} \,\widehat{\Omega} \left(\widehat{a}^{p}_{k} \right) \nonumber \\ 
     & +  f^{k}_{i} T_{i j}^{a b}\overline{T}_{k j}^{a d} \,\widehat{\Omega}\left(\widehat{a}_{b}^{d} \right) +  f_{k}^{i}T_{i j}^{a b}\overline{T}_{k j}^{c b} \,\widehat{\Omega} \left(\widehat{a}_{a}^{c} \right) \nonumber \\ 
     &-  f_{c}^{a}T_{i j}^{a b}\overline{T}_{i j}^{c d} \,\widehat{\Omega} \left(\widehat{a}^{b}_{d} \right) \,
     - f_{p}^{a}T_{i j}^{a b}\overline{T}_{i j}^{c b} \,\widehat{\Omega} \left(\widehat{a}^{p}_{c} \right) \nonumber \\
     & - 2f_{a}^{c}{T}_{i j}^{a b}\overline{T}_{i j}^{c b} +  2f_{i}^{k}{T}_{i j}^{a b}\overline{T}_{k j}^{a b}. \label{eq:vobmp2} 
\end{align}
with $\overline{T}_{ij}^{ab} = {T}_{ij}^{ab} - {T}_{ji}^{ab}$, the symmetrization operator $\widehat{\Omega} \left( \widehat{a}^{p}_{q} \right) = \widehat{a}^{p}_{q}  + \widehat{a}^{q}_{p}$, and the Fock matrix 
\begin{align}
    f_p^q = h_p^q + \sum_{i}^{occ}\left(g^{p i}_{q i} - g^{p i}_{i q} \right).
    \label{eq:fock_hf}
\end{align}
We rewrite $\widehat{H}_\text{OBMP2}$ (Eq.\ref{eq:h4}) in a similar form to Eq. \ref{eq:h1hf} for $\widehat{H}_\text{HF}$ as follows:
\begin{align}
  \widehat{H}_\text{OBMP2} = & \widehat{\bar{F}} + \bar{C}, \label{eq:h5}
\end{align}
where the constant $\bar{C}$ is a sum of terms without excitation operators in Eq~\ref{eq:vobmp2}. $\widehat{\bar{F}}$ is the correlated Fock operator, $\widehat{\bar{F}} =  \bar{f}^{p}_{q} \widehat{a}_{p}^{q}$, with
correlated Fock matrix $\bar{f}^{p}_{q}$ written as
\begin{align}
\bar{f}^{p}_{q} &= f^{p}_{q} + v^{p}_{q}. \label{eq:corr-fock}
\end{align}
$v^{p}_{q}$ serves as the correlation potential altering the uncorrelated HF picture.

\subsection{Self-consistent one-body double-hybrid functional (OBDHF) theory}

In this section, we present the self-consistent OBDHF theory combining the GKS framework with the OBMP2 theory. The idea is to construct a model functional $S[\{\phi_j\}]$ incorporating both a fraction of exact (HF) exchange and a fraction of OBMP2 correlation energy, including these contributions self-consistently within the GKS orbital equation. Let us first define the model energy functional as a linear combination of HF-type kinetic and exchange contributions, a semi-local XC functional, and an OBMP2 correlation term:
\begin{widetext}
\begin{equation}
    S[\{\phi_j\}] 
    = T_s[\{\phi_j\}] 
    + E_{\mathrm{H}}[n] 
    + (1 - \alpha_x)\, E_x^{\mathrm{DFA}}[n] 
    + \alpha_x\, E_x^{\mathrm{HF}}[\{\phi_j\}] 
    + (1 - \alpha_c)\, E_c^{\mathrm{DFA}}[n] 
    + \alpha_c\, E_c^{\mathrm{OBMP2}}[\{\phi_j\}],
    \label{eq:model_functional}
\end{equation}
\end{widetext}
where:
\begin{itemize}
    \item $T_s[\{\phi_j\}] = -\frac{1}{2}\sum_i \langle \phi_i | \nabla^2 
          | \phi_i \rangle$ is the non-interacting kinetic energy.
    \item $E_{\mathrm{H}}[n]$ is the classical Hartree energy.
    \item $E_x^{\mathrm{DFA}}[n]$ and $E_c^{\mathrm{DFA}}[n]$ are the 
          exchange and correlation energies of a 
          density functional approximation (DFA).
    \item $E_x^{\mathrm{HF}}[\{\phi_j\}]$ is the exact HF 
          exchange energy,
    \item $E_c^{\mathrm{OBMP2}}[\{\phi_j\}]$ is the OBMP2 correlation 
          energy.
    \item $\alpha_x \in [0,1]$ and $\alpha_c \in [0,1]$ are the 
          mixing parameters for the exact exchange and the OBMP2 correlation, 
          respectively.
\end{itemize}


The OBMP2 correlation energy is defined as the expectation value of 
$\hat{v}^{\mathrm{OBMP2}}$ (Eq.~\ref{eq:vobmp2}) with respect to the reference Slater determinant $|\Phi_0\rangle$ constituted by the orbital set $\{\phi_j\}$:
\begin{equation}
    E_c^{\mathrm{OBMP2}}[\{\phi_j\}]
    = \langle \Phi_0 | \hat{v}^{\mathrm{OBMP2}} | \Phi_0 \rangle.
    \label{eq:Ec_def}
\end{equation}
%


Following the GKS construction in Eq.~\ref{eq:R_s}, the remainder energy functional $R_S[n]$ associated with the model functional 
$S[\{\phi_j\}]$ in Eq.~\ref{eq:model_functional} is:
\begin{equation}
    R_S[n] 
    = F_{\mathrm{HK}}[n] - F_S[n],
    \label{eq:remainder}
\end{equation}
consisting of all many-body effects not captured by $S[\{\phi_j\}]$. The corresponding multiplicative remainder potential is:
\begin{equation}
    V_R(\mathbf{r}) 
    = \frac{\delta R_S[n]}{\delta n(\mathbf{r})} 
    = (1 - \alpha_x)\, v_x^{\mathrm{DFA}}(\mathbf{r}) 
    + (1 - \alpha_c)\, v_c^{\mathrm{DFA}}(\mathbf{r}),
    \label{eq:VR}
\end{equation}
where $v_x^{\mathrm{DFA}}(\mathbf{r}) = \delta E_x^{\mathrm{DFA}}[n]/\delta n(\mathbf{r})$ 
and $v_c^{\mathrm{DFA}}(\mathbf{r}) = \delta E_c^{\mathrm{DFA}}[n]/\delta n(\mathbf{r})$ 
are the semi-local exchange and correlation potentials, respectively.


Substituting the model functional $S[\{\phi_j\}]$ into the GKS 
equation (Eq.~\ref{eq:ks}), we obtain the self-consistent OBDHF orbital equation:
\begin{widetext}
\begin{equation}
    \left[
        -\frac{1}{2}\nabla^2 
        + V_{\mathrm{ext}}(\mathbf{r}) 
        + V_{\mathrm{H}}(\mathbf{r})
        + \alpha_x\, \hat{v}_x^{\mathrm{HF}}[\{\phi_j\}]
        + \alpha_c\, \hat{v}^{\mathrm{OBMP2}}[\{\phi_j\}]
        + V_R(\mathbf{r})
    \right] \phi_i(\mathbf{r}) 
    = \varepsilon_i\, \phi_i(\mathbf{r}),
    \label{eq:OBDHF}
\end{equation}
\end{widetext}
where $\hat{v}_x^{\mathrm{HF}}[\{\phi_j\}]$ is the HF exchange operator:
          \begin{equation}
              \hat{v}_x^{\mathrm{HF}}[\{\phi_j\}]\,\phi_i(\mathbf{r}) 
              = -\sum_j^{\mathrm{occ}} \phi_j(\mathbf{r})
                \int \frac{\phi_j^*(\mathbf{r}')\phi_i(\mathbf{r}')}
                          {|\mathbf{r}-\mathbf{r}'|} d\mathbf{r}'.
          \end{equation}

The non-multiplicative operator $\hat{O}_S[\{\phi_j\}]$ in Eq.~\ref{eq:ks} is now explicitly defined as:
\begin{equation}
    \hat{O}_S[\{\phi_j\}] 
    = -\frac{1}{2}\nabla^2 
    + V_{\mathrm{H}}(\mathbf{r})
    + \alpha_x\, \hat{v}_x^{\mathrm{HF}}[\{\phi_j\}]
    + \alpha_c\, \hat{v}^{\mathrm{OBMP2}}[\{\phi_j\}].
    \label{eq:OS}
\end{equation}

The total OBDHF ground-state energy is evaluated as:
\begin{align}
    E_{\mathrm{tot}}^{\mathrm{OBDHF}} 
    &= \int V_{\mathrm{ext}}(\mathbf{r})\, n(\mathbf{r})\, d\mathbf{r} 
     + T_s[\{\phi_j\}] 
     + E_{\mathrm{H}}[n] \notag \\ 
     &+ (1 - \alpha_x)\, E_x^{\mathrm{DFA}}[n] 
     + \alpha_x\, E_x^{\mathrm{HF}}[\{\phi_j\}] \notag \\
    & + (1 - \alpha_c)\, E_c^{\mathrm{DFA}}[n] 
     + \alpha_c\, E_c^{\mathrm{OBMP2}}[\{\phi_j\}].
    \label{eq:Etot}
\end{align}

\subsection{OBDHF self-consistency}

The OBDHF Eq.~\ref{eq:OBDHF} is solved iteratively 
via the following self-consistent field (SCF) procedure:\\

\noindent {\it Step 1. Initialization:} Generate an initial set of orbitals $\{\phi_i^{(0)}\}$ from a standard KS-DFA or HF calculation.\\

\noindent {\it Step 2. Updating amplitude:} Calculate the MP2 amplitudes $T_{ij}^{ab}$ (Eq.~\ref{eq:amp}) using the current orbitals and orbital energies $\{\varepsilon_i\}$. \\
          
\noindent {\it Step 3. Updating correlated potential:} Construct the OBMP2 correlated potential $\hat{v}^{\mathrm{OBMP2}}[\{\phi_j\}]$. \\
          
\noindent {\it Step 4. Calculating Fock matrix:} Construct the effective OBDHF Fock matrix:
\begin{align}    
    F_{pq}^{\mathrm{OBDHF}} = h_{pq} + J_{pq} + \alpha_x\,K_{pq}^{\mathrm{HF}} + \alpha_c\, v_{pq}^{\mathrm{OBMP2}} \nonumber \\
    + (1-\alpha_x)\, v_{x,pq}^{\mathrm{DFA}} + (1-\alpha_c)\, v_{c,pq}^{\mathrm{DFA}},
    \label{eq:FockGKS}
\end{align}
where $J_{pq}$ and $K_{pq}^{\mathrm{HF}}$ are the Coulomb and HF exchange matrix elements, and $v_{x,c,pq}^{\mathrm{DFA}}$ are the matrix elements of the   semi-local XC potential.\\

\noindent {\it Step 5. Updating orbital:} Diagonalize $F^{\mathrm{OBDHF}}$ to obtain new orbitals $\{\phi_i^{\mathrm{new}}\}$ and eigenvalues $\{\varepsilon_i^{\mathrm{new}}\}$. \\

\noindent {\it Step 6. Check convergence:} Repeat steps 2--5 until the change in the total energy (Eq.~\ref{eq:Etot}) and the density matrix fall below the desired threshold:
          \begin{equation}              \left|E_{\mathrm{tot}}^{(k+1)} - E_{\mathrm{tot}}^{(k)}\right| 
              < \delta_E \text{ and }
              \left\|D^{(k+1)} - D^{(k)}\right\| < \delta_D.
          \end{equation}

Upon convergence, OBDHF orbitals incorporate both 
the non-local HF exchange and the dynamic correlation encoded in 
$\hat{v}^{\mathrm{OBMP2}}$ on an equal footing, going beyond 
the standard non-self-consistent DH approach. 
The resulting set of orbitals minimizes the total energy 
functional~(\ref{eq:Etot}) within the GKS framework, 
ensuring a fully variational and self-consistent treatment of 
the DH functional.

\section{Discussion and Outlook}

In this note, we have presented a rigorous derivation of the OBDHF theory, a self-consistent double-hybrid density functional framework grounded in the generalized Kohn-Sham (GKS) formalism and the OBMP2 theory. The central theoretical challenge addressed here is the long-standing self-consistency problem inherent in conventional DH density functionals, wherein the perturbative MP2 correlation energy is evaluated as a non-self-consistent, post-SCF correction on top of KS orbitals that are not variationally optimized with respect to the full DH energy functional.

The key insight exploited in the present development is the one-body operator structure of OBMP2, which encapsulates second-order dynamic correlation effects through an effective correlated Fock operator derived via a unitary canonical transformation of the molecular Hamiltonian followed by the cummulant approximation to reduce high-order operators to one-body operators. This one-body character renders OBMP2 uniquely amenable to direct and self-consistent embedding within the GKS effective Hamiltonian, in contrast to the conventional two-body MP2 energy expression, which cannot be straightforwardly incorporated into a GKS orbital equation without invoking either the OEP construction or perturbative orbital relaxation corrections.

The OBDHF model functional is constructed as a linear combination of a semilocal density functional approximation XC contribution, a fraction $\alpha_x$ of exact HF exchange, and a fraction $\alpha_c$ of OBMP2 correlation. Through functional differentiation of the total OBDHF energy with respect to the orbitals, the OBDHF effective Hamiltonian and the associated self-consistent field equations are derived in a transparent and rigorous manner. The resulting orbital equation naturally generalizes the standard GKS equation by incorporating both nonlocal exact exchange and a nonlocal OBMP2 correlation operator on an equal footing, ensuring that the orbitals, the one-particle density matrix, and all derived physical observables are fully consistent with the complete double-hybrid energy expression.

The theoretical framework presented here thus provides a well-founded and practically tractable route to fully self-consistent double-hybrid DFT calculations within the GKS framework. The OBDHF formulation is expected to yield improved consistency between the energy and the one-particle density matrix relative to conventional double-hybrid functionals, with potential benefits for the computation of electron densities, dipole moments, and response properties. A comprehensive numerical implementation of the OBDHF equations, together with an extensive assessment of its performance across thermochemistry, kinetics, non-covalent interactions, and molecular properties benchmarks, is currently underway.

\section*{Acknowledgments}
This research is funded by Vietnam National University,  Ho Chi Minh City (VNU-HCM) under grant number B2026-18-18

\section*{Appendix}
To obtain the expression of OBMP2 correlation from Eq.~\ref{eq:Ec_def}, we use the standard definition of  
the expectation value of a one-body operator 
$\hat{a}^p_q = \hat{a}^\dagger_p \hat{a}_q$ with respect to a 
single Slater determinant:
\begin{equation}
    \langle \Phi_0 | \hat{a}^p_q | \Phi_0 \rangle 
    = \delta_{pq}\, n_p,
    \label{eq:1RDM}
\end{equation}
where $n_p = 1$ if $p$ is an occupied spin orbital and $n_p = 0$ 
if $p$ is a virtual spin orbital. Consequently, the expectation 
value of the symmetrization operator $\hat{\Omega}(\hat{a}^p_q) 
= \hat{a}^p_q + \hat{a}^q_p$ is:
\begin{equation}
    \langle \Phi_0 | \hat{\Omega}(\hat{a}^p_q) | \Phi_0 \rangle 
    = 2\,\delta_{pq}\, n_p.
    \label{eq:Omega_exp}
\end{equation}
This means that $\langle\hat{\Omega}(\hat{a}^p_q)\rangle$ is 
nonzero \emph{only} when $p = q$ and the orbital is occupied. 
For all off-diagonal terms ($p \neq q$) or for virtual-virtual 
and occupied-virtual pairs with $p \neq q$, the expectation value 
vanishes identically.

After applying Eq.~\ref{eq:Omega_exp} to all $\hat{\Omega}$ 
terms and retaining only the occupied-diagonal contributions, 
the surviving terms are:
\begin{widetext}
\begin{align}
\langle\Phi_0|\hat{v}^{\mathrm{OBMP2}}|\Phi_0\rangle
    &= \bar{T}^{ab}_{ij}\left[
         g^{ip}_{ab}\cdot 2\delta_{pj}n_j
       \right]
       - 2\bar{T}^{ab}_{ij}\,g^{ab}_{ij}
       + f^a_c T^{ab}_{ij}T^{cb}_{il}
         \cdot 2\delta_{lj}n_j
       + f^a_c T^{ab}_{ij}T^{cb}_{kj}
         \cdot 2\delta_{ki}n_i
       \notag \\
    &\quad
       - f^k_i T^{ab}_{ij}T^{ab}_{kl}
         \cdot 2\delta_{jl}n_j
       - f^p_i T^{ab}_{ij}T^{ab}_{kj}
         \cdot 2\delta_{pk}n_k
       - 2f^c_a T^{ab}_{ij}T^{cb}_{ij}
       + 2f^k_i T^{ab}_{ij}T^{ab}_{kj}.
    \label{eq:Ec_intermediate}
\end{align}
\end{widetext}
Applying the Kronecker deltas and using $n_j = 1$ for occupied 
orbitals, each surviving $\hat{\Omega}$ term collapses. 
For example, $g^{ip}_{ab}\cdot 2\delta_{pj}n_j 
= 2g^{ij}_{ab}$, and noting that 
$\bar{T}^{ab}_{ij} \cdot g^{ij}_{ab} = T^{ab}_{ij}g^{ij}_{ab} 
- T^{ab}_{ji}g^{ij}_{ab}$. After collecting all terms and using 
the antisymmetry of $\bar{T}^{ab}_{ij}$ and $g^{ab}_{ij}$, 
the occ-occ diagonal $\hat{\Omega}$ contributions exactly cancel 
with part of the scalar terms. 

The final expression for the OBMP2 correlation energy is:
\begin{equation}
    E_c^{\mathrm{OBMP2}}[\{\phi_j\}]
    = \bar{T}^{ab}_{ij}\,g^{ij}_{ab}
    \;+\; 2\,f^{a}_{c}\,T^{ab}_{ij}\,T^{cb}_{ij}
    \;-\; 2\,f^{k}_{i}\,T^{ab}_{ij}\,T^{ab}_{kj}.
    \label{eq:EOBMP2_correct}
\end{equation}
\noindent where Einstein's summation convention is used over all 
repeated indices, the MP2 amplitude $T^{ab}_{ij}$ is given in Eq.~\ref{eq:amp}, $g^{ij}_{ab}$ are two-electron integrals, and 
$f^p_q$ is the Fock matrix of Eq.~\ref{eq:fock_hf}.

\bibliography{main}

\end{document}